\documentclass[prb,twocolumn,10pt, superscriptaddress]{revtex4}
\usepackage{amsmath}
\usepackage{graphicx}
\begin{document}
\title{A valley-spin qubit in a carbon nanotube}
\author{E. A. Laird}
\affiliation{Kavli Institute of Nanoscience, Delft University of Technology, 2600 GA, Delft, The Netherlands.}
\author{F. Pei}
\affiliation{Kavli Institute of Nanoscience, Delft University of Technology, 2600 GA, Delft, The Netherlands.}
\author{L. P. Kouwenhoven}
\affiliation{Kavli Institute of Nanoscience, Delft University of Technology, 2600 GA, Delft, The Netherlands.}

\maketitle

\begin{figure*}
\vspace{-0.1cm}
\includegraphics[width=17cm]{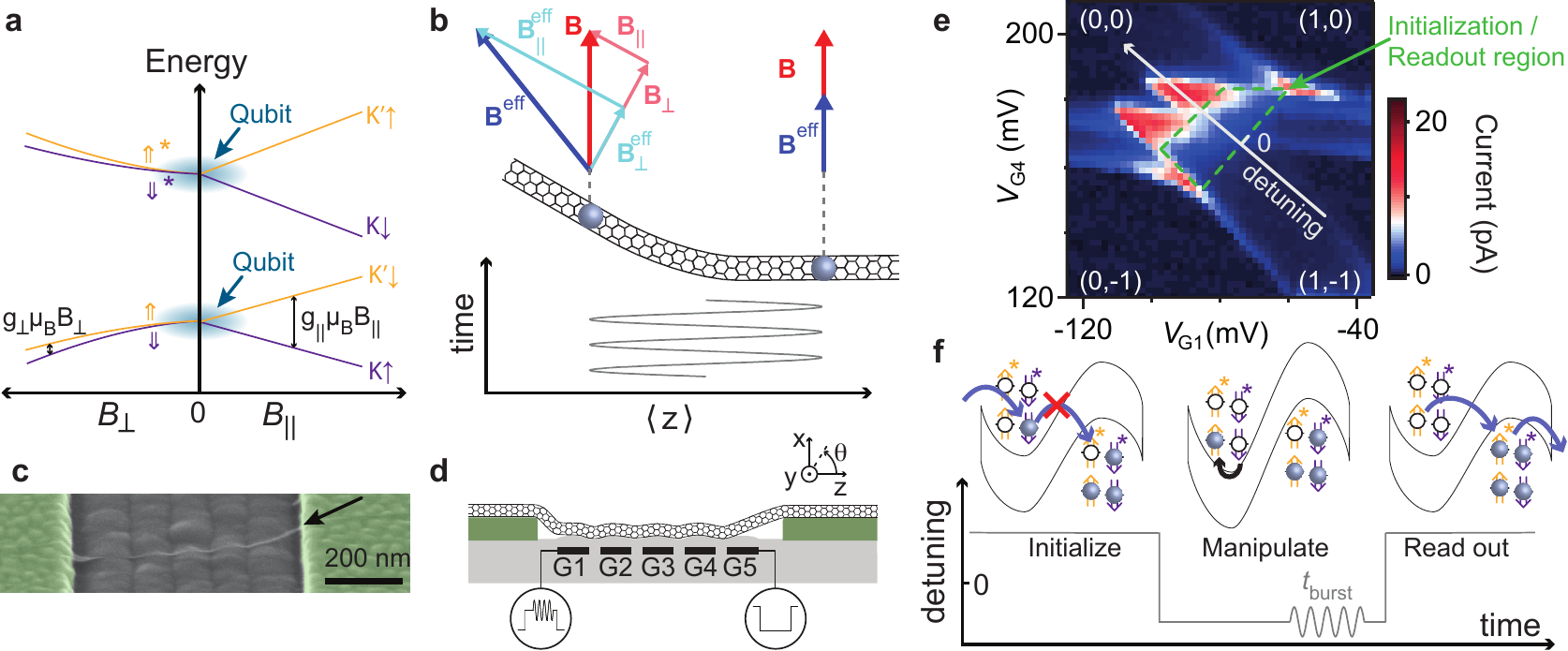}
\vspace{-0.3cm}
\caption{\footnotesize{{\bf Valley-spin resonance in a bent nanotube. a} Single-particle energy levels of a straight nanotube, as a function of magnetic field applied parallel or perpendicular to the nanotube. The two doublets are split with effective $g$-factors depending on the field orientation. Either doublet can act as a valley-spin qubit. For large $B_{||}$, the four states take the simple forms labelled on the right.  {\bf b}, Using a bend to mediate spin resonance. On the right, the nanotube is oriented perpendicular to ${\bf B}$, aligning $\bf{B}$ with a principal axis of the $\bf{g}$-tensor so that ${\bf B}^\mathrm{eff}$ is parallel to ${\bf B}$. However, on the left, ${\bf B}^\mathrm{eff}$ is a sum of parallel ($B^\mathrm{eff}_{||}$) and perpendicular ($B^\mathrm{eff}_\perp$) components, which are related to $B_{||}$ and $B_\perp$ by different components of $\bf{g}$. An electron driven back and forth across the bend experiences $\mathbf{B}^\mathrm{eff}$ to vary periodically in direction, leading to spin resonance. {\bf c}, SEM micrograph and {\bf d}, schematic of the measured device. The bent nanotube (marked by arrow) spans a trench between source and drain contacts (green). Voltages applied to gates G1-5 beneath the trench control a double quantum dot potential and allow for microwave driving and pulsed control. The coordinates $x$, $z$ and $\theta$ are defined in {\bf d}. {\bf e}, Current through the device as a function of gate voltages on G1 and G4 close to the (1,-1)$\rightarrow$(0,0) transition, with 5~mV source-drain bias. Double-dot occupations in the four adjoining regions of gate space are indicated, and the detuning axis is marked by a grey arrow. Inside the green quadrilateral, Pauli blockade makes the current sensitive to the valley-spin state, an effect that is exploited for qubit initialization and readout. {\bf f}, Cycle of gate voltage pulses for qubit manipulation. Beginning in Pauli blockade initializes a long-lived blocked two-qubit state, for example $\Downarrow \Downarrow^*$. The device is then configured in (1,-1) and a microwave burst of duration $t_\mathrm{burst}$ is applied, which on resonance flips one of the qubits, leading to a state such as $\Uparrow \Downarrow^*$. For readout, the device is returned to the Pauli blockade configuration, and if a qubit flip has occurred in either dot, an electron tunnels out into the right lead. The average current is proportional to  the probability of a qubit flip during the manipulation stage. }}
\vspace{-0.5cm}
\end{figure*}

Although electron spins in III-V semiconductor quantum dots have shown great promise as qubits~\cite{Hanson:2007p597,Nowack:2007p150, NadjPerge:2010p982}, a major challenge is the unavoidable hyperfine decoherence in these materials. In group IV semiconductors, the dominant nuclear species are spinless, allowing for qubit coherence times~\cite{Balasubramanian:2009ub,Tyryshkin:2011uz,Pla:2012jj} up to 2~s. Carbon nanotubes are a particularly attractive host material, because the spin-orbit interaction with the valley degree of freedom allows for electrical manipulation of the qubit~\cite{Kuemmeth:2008p676, Bulaev:2008p566,  Flensberg:2010p885, Palyi:2011p1007, Klinovaja:2011ht}. In this work, we realise such a qubit in a nanotube double quantum dot~\cite{Churchill:2009p712, Churchill:2009p711}. The qubit is encoded in two valley-spin states, with coherent manipulation via electrically driven spin resonance~\cite{Nowack:2007p150, NadjPerge:2010p982} (EDSR) mediated by a bend in the nanotube. Readout is performed by measuring the current in Pauli blockade~\cite{Pei:2012vz}. Arbitrary qubit rotations are demonstrated, and the coherence time is measured via Hahn echo. Although the measured decoherence time is only 65~ns in our current device, this work offers the possibility of creating a qubit for which hyperfine interaction can be virtually eliminated.

The operating principle of the qubit~\cite{Flensberg:2010p885} is explained in Fig 1a-b. The scheme relies on the existence of the valley degree of freedom, which classifies electron states according to the orbital magnetic moment around the nanotube~\cite{Minot:2004tc}. Spin ($\uparrow, \downarrow$) and valley ($K,K'$) quantum numbers together lead to four electron states, which are separated by spin-orbit coupling~\cite{Kuemmeth:2008p676,Bulaev:2008p566,Jhang:2010do,Jespersen:2011p1006} into two doublets, denoted \{$\Uparrow, \Downarrow$\} and \{$\Uparrow^*, \Downarrow^*$\}. The resulting single-particle spectrum is plotted in Fig.~1a as a function of magnetic field $\mathbf{B}$ applied either parallel ($B_{||}$) or perpendicular ($B_{\perp}$) to the nanotube axis. Whereas $B_{||}$ couples to both spin and valley magnetic moments, $B_{\perp}$ couples only via the spin, so that the energy splitting within each doublet  depends on field direction. This is parameterized by an anisotropic effective $\mathbf{g}$-tensor with components $g_{||}$ for parallel and $g_\perp < g_{||}$ for perpendicular field direction. Although for large $B_{||}$, the four states \{$\Uparrow, \Downarrow, \Uparrow^*, \Downarrow^*$\} take the simple forms labelled on the right of Fig. 1a, in general they are entangled combinations of spin and valley eigenstates. Either doublet can act as a qubit, depending on the quantum dot occupation.

Qubit manupilation using an electric field is possible if the nanotube contains a bend (Fig. 1b)~\cite{Flensberg:2010p885}. This can be understood by considering each doublet as an effective spin-1/2 system. The energy splitting can then be considered as due to isotropic Zeeman coupling to an effective magnetic field $\mathbf{B}_\mathrm{eff}=\mathbf{g \cdot B}/g_\mathrm{s}$, where $g_\mathrm{s}\approx 2$ is the Zeeman $g$-factor. Where the nanotube axis is perpendicular to $\mathbf{B}$ (right side of Fig.~1b),  $\mathbf{B}_\mathrm{eff}$ and $\mathbf{B}$ are parallel. However, where the nanotube axis is at an angle (left side of Fig.~1b), the components $B_{||}$ and $B_\perp$ contribute differently to $\mathbf{B}_\mathrm{eff}$, tilting $\mathbf{B}_\mathrm{eff}$ away from $\mathbf{B}$. Under application of an ac electric field, an electron driven back and forth across the bend experiences a $\mathbf{B}_\mathrm{eff}$ that oscillates in direction. With the driving frequency set to $f =\Delta E /h$, where $\Delta E=g_\mathrm{s}\mu_\mathrm{B}|\mathbf{B}_\mathrm{eff}|$ is the qubit energy splitting, $h$ is Planck's constant, and $\mu_\mathrm{B}$ is the Bohr magneton, this drives resonant transitions between the qubit states, allowing for arbitrary coherent single-qubit operations. Because the two qubit states do not have the same spin, driving transitions between them in this way leads to EDSR.

As in previous experiments~\cite{Hanson:2007p597}, the qubit is realized in a double quantum dot where it can be initialized and read out by exploiting Pauli blockade. The measured device (Fig. 1c) consists of a single electrically contacted nanotube which is bent by touching the substrate (Fig. 1d)~\cite{Pei:2012vz}. The electrical potential is controlled using voltages applied to nearby gate electrodes, which were tuned to configure the double quantum dot close to the (1,-1)$\rightarrow$(0,0) transition, where numbers in brackets denote the occupation of left and right dots, with positive (negative) numbers indicating electrons (holes)~\cite{Steele:2009p716}. 

Figure 1e shows the current with 5~mV source-drain bias across the device, measured as a function of gate voltages coupled to the left and right quantum dots without pulses or microwaves applied. The dashed quadrilateral outlines the region of gate space where Pauli blockade strongly suppresses the current, with the corresponding valley-spin energy levels shown in the first panel of Fig. 1f. Although electron tunneling through the double dot is energetically allowed, it is suppressed by selection rules on valley and spin~\cite{Pei:2012vz} because an electron loaded from the left in the $\Uparrow$ or $\Downarrow$ state is forbidden by Pauli exclusion from entering the corresponding filled state in the right dot. Pauli blockade is broken by tunneling events that do not conserve spin and valley, for example due to spin-orbit coupling combined with disorder~\cite{Palyi:2010fs, Pei:2012vz}, which give rise to the leakage current near the tips of the triangles in Fig.~1e. 

The combined two-qubit state is defined by the states of the unpaired electrons in left and right dots. Although Pauli blockade applies for any combination of $\{\Uparrow, \Downarrow\}$ on the left and $\{\Uparrow^*, \Downarrow^*\}$ on the right, the rate at which it is broken by disorder-induced valley mixing is different for different states because they contain different superpositions of valley-spin quantum numbers (see Supplementary)~\cite{Reynoso:2012ul,Pei:2012vz}.  The leakage current is therefore sensitive to the rate at which qubits are flipped.

To detect EDSR, the current is measured with the following cycle of pulses and microwave bursts applied to the gates (Fig. 1f)~\cite{Hanson:2007p597}. Beginning with the double dot configured inside the quadrilateral marked in Fig.~1e initializes a long-lived blocked state. The device is then pulsed along the detuning axis defined in Fig. 1e to a configuration in (1,-1), during which the qubits are manipulated by applying microwaves for a time $t_\mathrm{burst}$. If the microwave drive is on resonance with either qubit, it will drive coherent rotations. For readout, the device is returned to the starting configuration. If a qubit flip has occurred, so that the device is no longer in a long-lived state, an electron will tunnel through the device. Averaged over many cycles, the current is proportional to the qubit flip probability during the manipulation stage.

\begin{figure*}
\vspace{-0.5cm}
\includegraphics[width=17cm]{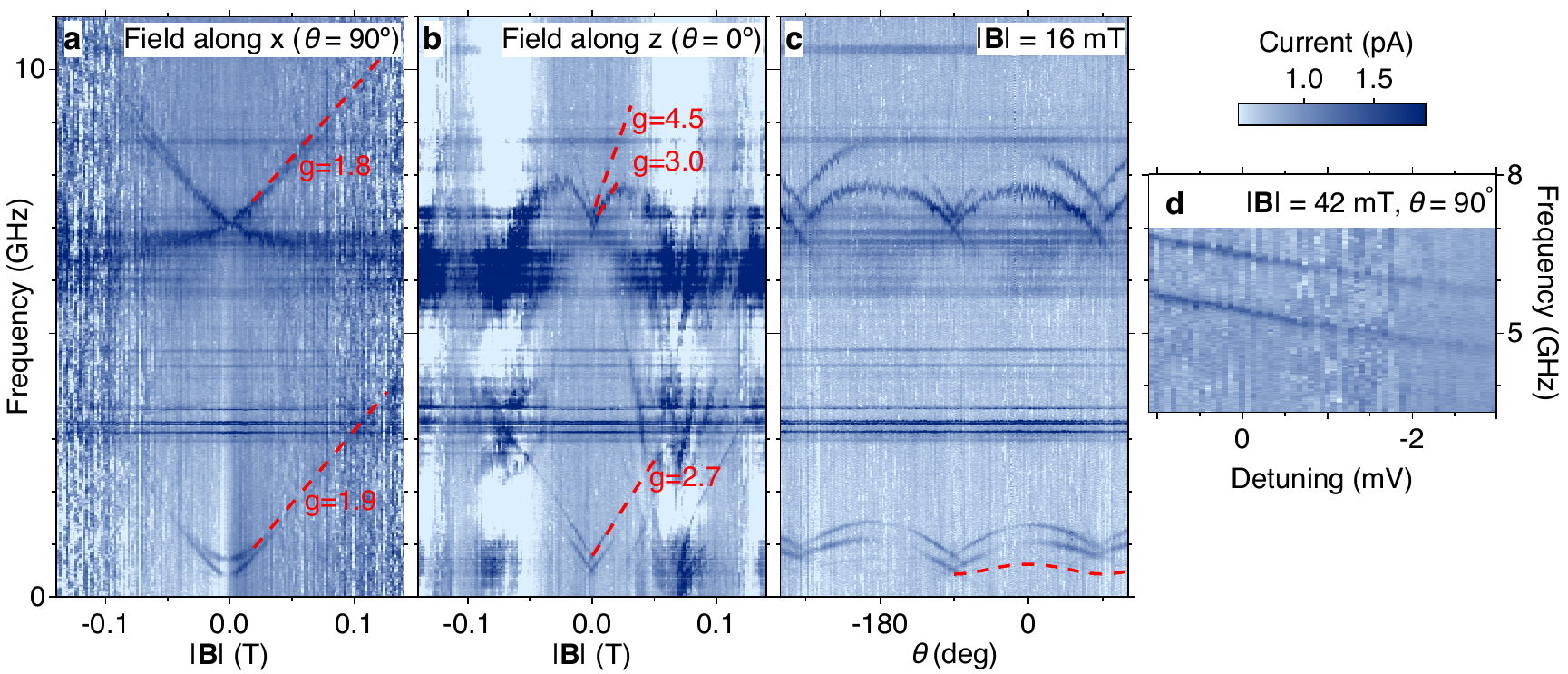}
\vspace{-0.4cm}
\caption{\footnotesize{{\bf Spin resonance spectroscopy.} Average current with the pulse cycle of Fig.~1f applied,  measured as a function of microwave frequency and {\bf a}, magnetic field along the $x$ axis, {\bf b}, magnetic field along the $z$ axis, {\bf c}, magnetic field angle relative to the $z$-axis, and {\bf d},~detuning during the manipulation pulse. Measured $g$-factors for selected resonance lines are indicated in {\bf a} and {\bf b}. Taking perpendicular and parallel $g$-factors for the lowest resonance predicts~\cite{Flensberg:2010p885} the expected resonance frequency marked in {\bf c}, which does not agree well with the observed spectrum. To make the signal clearer, a frequency-independent background has been subtracted from all four plots. The data in {\bf d} was taken for slightly different device tuning than in {\bf a-c}.}}
\vspace{-0.5cm}
\end{figure*}

\begin{figure}
\vspace{0.3cm}
\includegraphics[width=8.8cm]{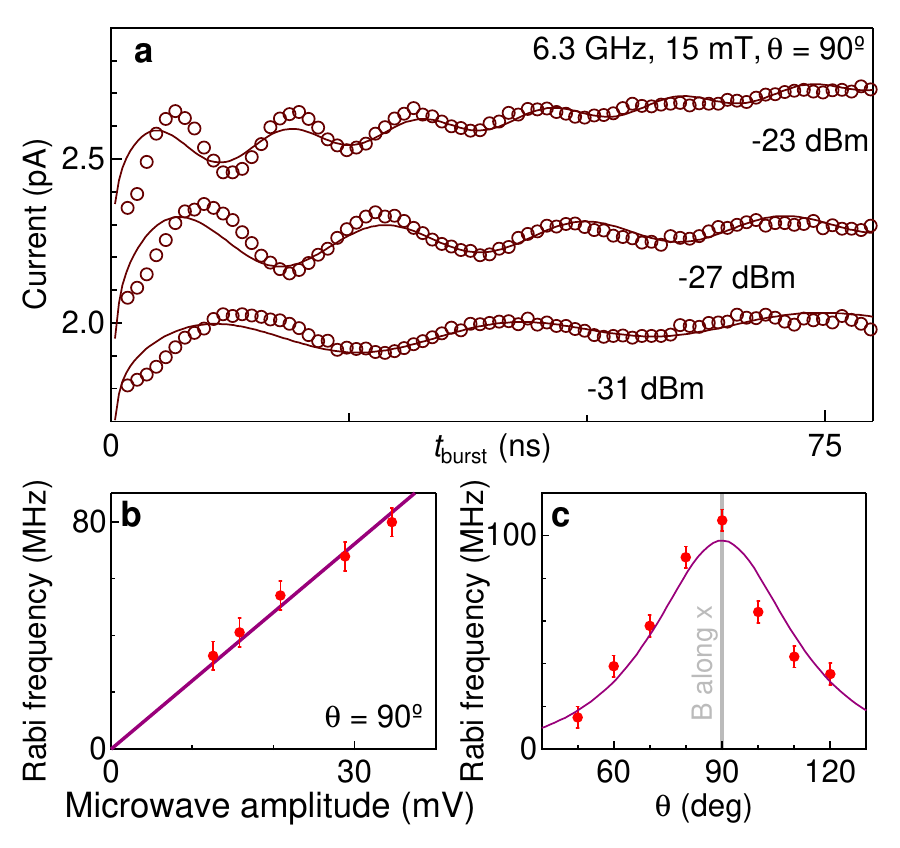}
\vspace{-0.5cm}
\caption{\footnotesize{{\bf Coherent qubit manipulation a}, Points: Current as a function of $t_\mathrm{burst}$ measured on the highest resonance, for three different values of the microwave power at the sample holder. Rabi oscillations show coherent control of the qubit. Lines: Fits to the function $I_0 + a t_\mathrm{burst}^{-0.5}\cos (2\pi f_\mathrm{R} t_\mathrm{burst} + \pi/4) + bt_\mathrm{burst}$, appropriate in the limit of weak driving and slow dephasing~\cite{Koppens:2007p639}.  $I_0$ is a constant background current, $f_\mathrm{R} $ is the Rabi frequency, and the $bt_\mathrm{burst}$ term parameterizes a weak ($<230$~fA) contribution from photon-assisted tunneling. Middle and upper traces are offset for clarity. (The device tuning changed slightly between Figs. 2 and 3, leading to a lower resonance frequency.) {\bf b}, Rabi frequency (points) as a function of rms microwave voltage at the sample holder. As expected, the data shows good agreement with a linear fit (line). {\bf c}, Rabi frequency (points) as a function of field angle at constant microwave power and frequency, showing a maximum for field along the $x$-axis. The data agrees well with a theoretical fit~\cite{Flensberg:2010p885} (line) taking the $g$-factors measured on the topmost lines in Fig. 2a and 2b, and treating the overall scaling as a fit parameter.}}
\vspace{-0.5cm}
\end{figure}

The measured EDSR spectrum is shown in Fig. 2 as a function of $B_{x}$, $B_{z}$, field angle $\theta$ in the $xz$ plane, and detuning. In each plot, EDSR is evident as an increased current at the resonance frequency, which depends on magnetic field. The spectrum is much more complex than previously measured for spin qubits~\cite{Hanson:2007p597} or expected from the level diagram in Fig.~1a, but nevertheless shows some features in agreement with theory~\cite{Flensberg:2010p885}. At low frequency (between $\sim$~0.8~GHz and $\sim4$~GHz), the spectrum exhibits an approximately linear increase of resonance frequency with $|\mathbf{B}|$ (Fig. 2a,b), with $g$-factors (dashed lines in Fig.~2a,b) higher for fields along the $z$ than along $x$, indicating coupling to the valley degree of freedom. However, the anisotropy is much less than expected from Coulomb blockade spectroscopy, which yields $g$-factors larger in the $z$ direction and a smaller in the $x$-direction~(see Supplementary). Furthermore, the angular dependence of the resonance frequency agrees only qualitatively with the prediction (dashed line in Fig. 2c) of Ref.~\cite{Flensberg:2010p885} given the measured $g$-factors along $x$ and $z$. This reflects the fact that the resonance lines in Fig.~2a,b do not extrapolate to zero, with the zero-field resonant frequency being larger in Fig.~2b.

A qualitatively unexpected feature is the pronounced series of resonances centred around 7~GHz. Such a high-frequency manifold could be expected from transitions between starred and unstarred states in Fig.~1a; however, the corresponding resonance frequency, set by the spin-orbit energy, would be $\sim$200~GHz, far higher than measured. We suggest that the complex spectrum instead reflects inter-dot exchange~\cite{Nowak}, the multiple bends in the nanotube induced by the surface, and possibly Rashba-type spin-orbit coupling~\cite{Klinovaja:2011ht}. The upper resonance is observed to shift as a function of detuning (Fig.~2d). We tentatively ascribe this to a a change of exchange with detuning, although it could also be due to shift of electron wavefunction around a bend. In either case, this effect could be useful to bring a qubit rapidly in and out of resonance with the microwave field by pulsing the detuning. Although spin-orbit mediated driving is expected~\cite{Nowack:2007p150, NadjPerge:2010p982, Flensberg:2010p885} to become more efficient at higher $|\mathbf{B}|$, the corresponding increase in resonant current is not observed. This is probably because the background leakage current increases with field, making qubit initialization less efficient.

Coherent operation of the qubit is demonstrated in Fig.~3 by measuring the resonant current as a function of $t_\mathrm{burst}$. The current is observed to oscillate (Fig. 3a), with the oscillation frequency proportional to the microwave driving amplitude (Fig.~3b). These Rabi oscillations arise from the coherent precession of the qubit state during the microwave burst. Although this data is measured for the highest-frequency resonance because it gave the best contrast, Rabi oscillations were also observed for the other resonance lines in Fig. 2a. 

\begin{figure}[h!]
\vspace{-0.2cm}
\includegraphics[width=8.8cm]{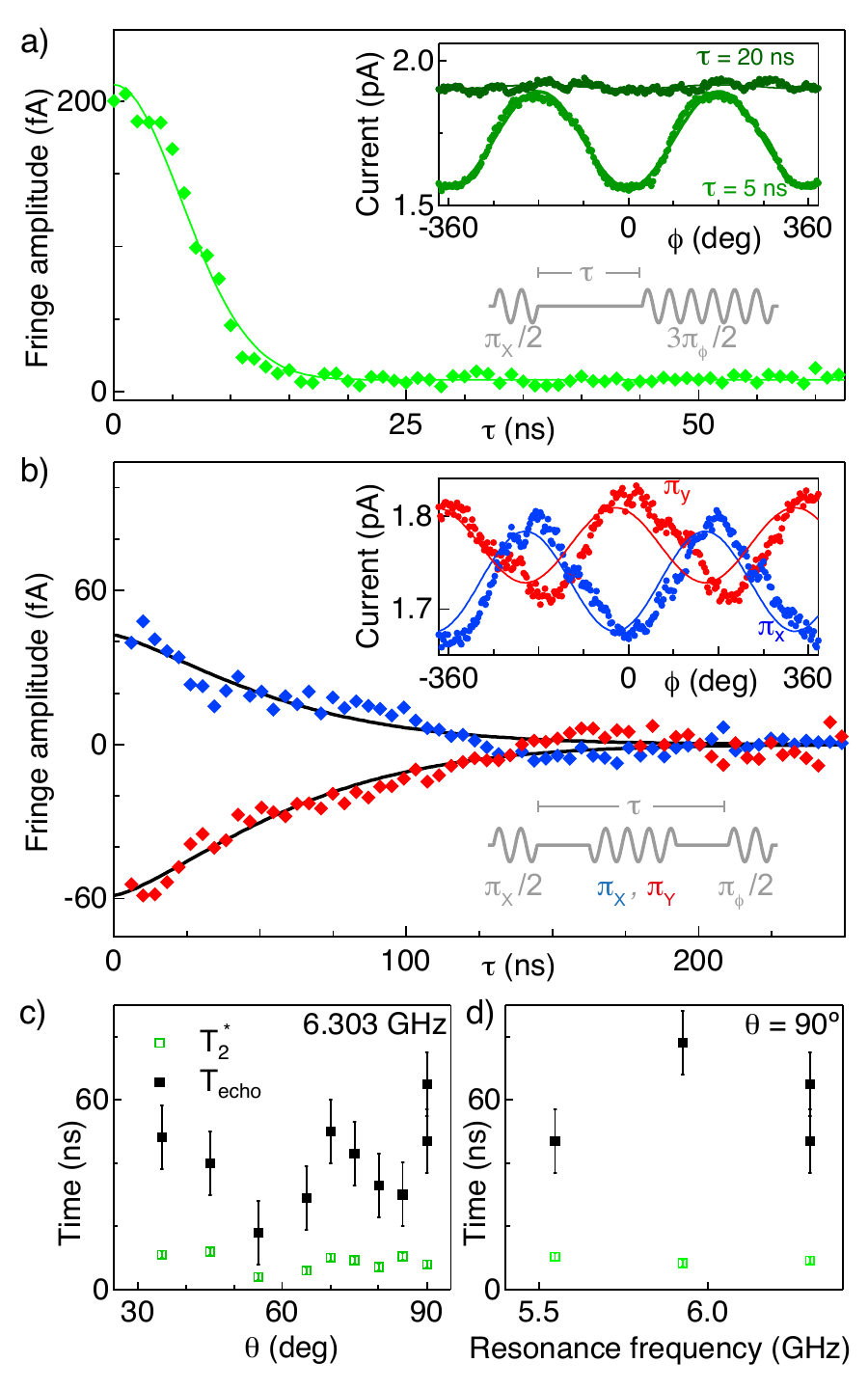}
\vspace{-0.5 cm}
\caption{\footnotesize{{\bf Universal control and measurement of coherence times. a}, Points: amplitude of Ramsey fringes, measured using a $\pi_x/2 - \tau - 3\pi_\phi/2$ microwave burst sequence under the same conditions as Fig. 3a, plotted as a function of $\tau$. Line: Fit to Gaussian decay, giving $T_2^* = 8 \pm1$ ns. Upper inset shows current (points) as a function of microwave phase for two values of $\tau$, together with cosinusoidal fits (lines). The observed oscillation for $\tau$ = 5 ns indicates that qubit rotations are achieved about arbitrary axes. {\bf b},~Extending the coherence time via echo. Main panel shows fringe amplitude (points) in an echo pulse sequence as a function of $\tau$, with the $\pi$ pulse phase chosen along $x$ (red) or $y$ (blue). Fits (lines) to a decay of the form $\mathrm{exp}(-(\tau/T_\mathrm{echo})^\alpha)$ with $\alpha=1.3 \pm 0.2$, give coherence time $T_\mathrm{echo} = 65 \pm 10$ ns. Upper inset: Current (points) as a function of $\pi/2$ phase for $\pi$ phase along $x$ and $y$, together with cosinusoidal fits (lines). As expected, the phase is reversed for $\pi$ rotations about orthogonal axes.  In both {\bf a} and {\bf b}, the lower inset shows a schematic of microwave burst sequence and definition of $\tau$. {\bf c}, Coherence times as a function of field angle, measured at constant frequency and Rabi frequency (and therefore increasing drive power for decreasing $\theta$.) {\bf d} Coherence times as a function of resonance frequency, measured at constant Rabi frequency with field applied along $x$. Within the experimental error, no significant dependence on angle or frequency is seen.}}
\end{figure}

Two other mechanisms have been proposed for coherent EDSR in nanotubes: a Rashba-like spin-orbit coupling induced by the electric field of the gates~\cite{Bulaev:2008p566, Klinovaja:2011ht}, and coupling to spatially inhomogeneous disorder~\cite{Palyi:2011p1007}. Measurement of the Rabi frequency as a function of field angle gives insight into the mechanism. As shown in Fig.~3c, the Rabi frequency at constant driving frequency and power is maximal with the field applied in the perpendicular direction. The dependence on angle agrees well with the bent nanotube prediction~\cite{Flensberg:2010p885}, taking the $g$-factors measured from Fig.~2a,b as fixed and treating the overall coupling as a free parameter. In contrast, for Rashba-like coupling, the spin-orbit field in our geometry is predominantly along $y$, making the Rabi frequency nearly independent of field angle in the $xz$ plane\cite{Klinovaja:2011ht}. For disorder-mediated EDSR, the expected angle depence of Rabi frequency is unknown\cite{Palyi:2011p1007}. The data in Fig.~3 is therefore consistent with the bend being the main mechanism for EDSR in this device, although other mechanisms probably contribute. Figure 2 provides evidence in particular of a Rashba-like contribution, in that the EDSR signal is evident even at at $|{\bf B}|=0$ (Figs. 2a,b), where bend-mediated EDSR is predicted to be absent~\cite{Flensberg:2010p885}. In contrast, the Rashba-like mechanism gives a finite signal even at zero field, as observed~\cite{Klinovaja:2011ht}.

While Rabi oscillations demonstrate qubit rotations about one axis of the Bloch sphere, for universal control, rotations about two independent axes are necessary. This is demonstrated by a Ramsey fringe experiment, in which the single microwave burst of Fig.~1f is replaced by a pair of bursts inducing qubit rotations by $\pi/2$ and $3\pi/2$ (Fig. 4a), with a phase difference $\phi$ between them. With the first burst taken as a rotation about the Bloch $X$ axis, the second burst induces a rotation about an axis offset by angle $\phi$ in the $XY$ plane. As expected, the current is found to oscillate as a function of $\phi$  depending whether the two rotations interfere constructively or destructively (Fig. 4a inset). As the burst interval $\tau$ is increased, the interference contrast decreases, yielding the dephasing time $T_2^*=8$~ns (Fig.~4a).

Coherence can be prolonged by inserting a $\pi$ burst into the sequence to cancel slowly varying dephasing sources via Hahn echo (Fig. 4b). With this sequence, the decoherence time from the decay of fringes is $T_\mathrm{echo}=65$~ns. Introduction of further $\pi$ bursts (Carr-Purcell decoupling) did not lead to a longer coherence time. The coherence time also does not depend significantly on field angle (Fig.~4c) or magnitude (Fig.~4d).

The measured $T_2^*$ is consistent with the hyperfine coupling measured previously in nanotube double dots, and therefore this may be the dominant dephasing mechanism (alternative mechanisms are considered in the Supplementary)~\cite{Hanson:2007p597,Churchill:2009p712, Churchill:2009p711,Reynoso:2011hm}. However, the rapid $T_\mathrm{echo}$ (compared with GaAs) that we observe would be inconsistent with hyperfine decoherence, unless the nuclear spin diffusion is extremely rapid~\cite{Hanson:2007p597}. We speculate that it is due to charge noise, which couples to the qubit because the resonance frequency depends on detuning~(Fig.~1d). Future experiments will measure nanotubes fabricated from isotopically purified $^{12}$C feedstock~\cite{Balasubramanian:2009ub} to definitively isolate the hyperfine contributions to qubit decoherence.

{\bf Methods}

The device was previously measured in a separate cooldown in Ref.~\cite{Pei:2012vz}, where the fabrication is described in detail. Measurements were performed at 270~mK in a $^3$He refrigerator equipped with a vector magnet. Except for Fig.~S1 of the Supplementary, the magnetic field was applied in the $xz$ plane defined in Fig.~1d, where the $x$-axis is normal to the chip and the $z$-axis runs perpendicular to the gates. Because the growth direction is uncontrolled, the nanotube is misaligned from this plane by approximately $6^\circ$. Schottky barriers with the contact electrodes defined the left and right barriers of the double quantum dot, while the central barrier was defined by an {\it np} junction. From the curvature of  charge transition lines in the stability diagram~\cite{VanDerWiel:2003p153}, the tunnel coupling is estimated as $t_\mathrm{c}=0.9 \pm 0.3$~meV. The double-dot occupancy was tuned using the gate voltages and determined by identifying the bandgap in the charge stability diagram, with Pauli blockade  recognized by a pronounced current suppression for one bias direction that was strongly affected by the magnetic field. The duration of the overall pulse cycle was typically 1~$\mu$s, split between a measurement pulse of $\sim$400~ns and a manipulation pulse of $\sim 600$~ns.
\vfill

\small
\bibliography{SpinQubitRefs}

\vskip 0.25in
\noindent
{\bf Acknowledgements}
We acknowledge G.A.~Steele, K.~ Flensberg, J. Klinovaja, D.~Loss, A.~P\'{a}lyi, J.~van~den~Berg, S.M.~Frolov, and V.S.~Pribiag for discussions. This work was supported by NWO/FOM.
\\ \\
{\bf Author Contributions}
F.P. fabricated and characterized the device. E.A.L. performed the experiment. All authors prepared the manuscript.
\\ \\
{\bf Additional Information}
Supplementary information accompanies this paper.
Correspondence and requests for materials should be addressed to E.A.L.~(e.a.laird@tudelft.nl).

\end{document}